\documentclass[aps,prl,10pt,twocolumn,superscriptaddress,amsmath,amssymb]{revtex4-2}
\usepackage{graphicx}   
\graphicspath{{Plots/}}
\usepackage{hyperref}   
\usepackage{siunitx}
\makeatletter
\let\saved@includegraphics\includegraphics
\AtBeginDocument{\let\includegraphics\saved@includegraphics}
\makeatother

\newcommand{\Dunit}{\si{\micro\metre\squared \, \second^{-1}}}
\newcommand{\Drunit}{\si{\radian\squared \, \second^{-1}}}
\newcommand{\Vunit}{\si{\micro\metre \,\second^{-1}}}

\usepackage{xcolor}
\definecolor{myBlue}{rgb}{0.1,0.1,0.7}

\begin{document}
\title{Delayed Active Swimmer in a Velocity Landscape}
\author{Viktor Holubec}
\email{viktor.holubec@matfyz.cuni.cz}
\affiliation{Department of Macromolecular Physics, Faculty of Mathematics and Physics,Charles University, 18000 Prague, Czech Republic}
\author{Alexander Fischer}
\affiliation{Molecular Nanophotonics Group, Peter Debye Institute for Soft Matter Physics, Universit\"at Leipzig, 04103 Leipzig, Germany}
\author{Giovanni Volpe}
\affiliation{Department of Physics, University of Gothenburg, 41296 Gothenburg, Sweden}
\author{Frank Cichos}
\email{cichos@physik.uni-leipzig.de}
\affiliation{Molecular Nanophotonics Group, Peter Debye Institute for Soft Matter Physics, Universit\"at Leipzig, 04103 Leipzig, Germany}

\date{\today}
\begin{abstract}
Self-propelled active particles exhibit delayed responses to environmental changes, modulating their propulsion speed through intrinsic sensing and feedback mechanisms. This adaptive behavior fundamentally determines their dynamics and self-organization in active matter systems, with implications for biological microswimmers and engineered microrobots.
Here, we investigate active Brownian particles whose propulsion speed is governed by spatially varying activity landscapes, incorporating a temporal delay between environmental sensing and speed adaptation. Through analytical solutions derived for both short-time and long-time delay regimes, we demonstrate that steady-state density and polarization profiles exhibit maxima at characteristic delays. Significantly, we observe that the polarization profile undergoes sign reversal when the swimming distance during the delay time exceeds the characteristic diffusion length, providing a novel mechanism for controlling particle transport without external fields. Our theoretical predictions, validated through experimental observations and numerical simulations, establish time delay as a crucial control parameter for particle transport and organization in active matter systems. These findings provide insights into how biological microorganisms might use response delays to gain navigation advantages and suggest design principles for synthetic microswimmers with programmable responses to heterogeneous environments.
\end{abstract}

\maketitle

\section{Introduction}
Active matter consists of particles operating far from equilibrium that convert energy into directed motion. These systems exhibit properties not possible in conventional materials that are constrained by thermal equilibrium symmetries, including persistent ring currents~\cite{Gnesotto2018BrokenDetailedBalance} and the breaking of mesoscopic action-reaction symmetry~\cite{Steffenoni2016InteractingBath, Basu2015StatForces}. Microscopic organisms, particularly bacteria~\cite{Cates2012DiffusivePhysics}, represent fascinating examples of active matter as they navigate heterogeneous landscapes created by spatial gradients of chemicals, nutrients, or light~\cite{Miller2001, VanHaastert2004, Raynaud2014}, as well as physical obstacles~\cite{Cates2015MIPS}. Understanding these systems is critical to advance our fundamental knowledge of non-equilibrium physics and to guide the development of synthetic microrobots for applications such as targeted medical interventions~\cite{Contera2019Nano}.

The distribution of active particles in heterogeneous environments has been studied under various simplifying assumptions. For microswimmers responding instantaneously to activity landscapes with negligible translational diffusion, their density $\rho$ and swim speed $v$ are inversely proportional, $\rho \sim 1/v$~\cite{Schnitzer1993}. However, real systems exhibit two critical features that complicate this picture: finite reaction times between sensing and response~\cite{Fukuoka2014,Friedrich2018}, and translational diffusion characterized by a coefficient $D$~\cite{Romanczuk2015OptimalCells,Diz-Munoz2016SteeringProtrusionsb,Landin2021,wang_spontaneous_2023}. Biological examples demonstrate the importance of temporal delays across scales: from gene expression affecting pattern formation~\cite{Gaffney2006}, to social amoebae using delays in intercellular signaling to coordinate collective behavior~\cite{Gregor2010} and maintain directional chemotaxis~\cite{Skoge2014}, to green algae tuning phototactic steering through delays matched to their rotational period~\cite{Leptos2023}. These examples illustrate how organisms have evolved to exploit temporal delays as control parameters rather than limitations.

Previous theoretical work has explored the effects of these parameters separately. For negligible translational diffusivity, Refs.~\cite{Volpe2016,Volpe2018} showed that the product of a short time delay $\tau$ and rotational diffusivity $D_\theta$ modifies the density profile as $\rho \sim 1/v^{1+D_\theta \tau}$, increasing particle accumulation in low-speed regions. Conversely, for negligible delay, Refs.~\cite{Sokker2021,Auschra2021,AuschraCurved2021} demonstrated that translational diffusion introduces a nonzero decay length of polarization profiles near activity steps and substantially modifies density distributions. Despite these advances, a unified framework accounting for both finite delays and nonzero translational diffusion has remained elusive.

In this letter, we derive analytical expressions for density and polarization profiles incorporating both nonzero translational diffusivity $D$ and finite delay times $\tau$. Our theoretical model reveals three key phenomena: (1) short delays enhance density maxima in low-activity regions; (2) polarization profiles undergo sign inversion when swimming distance during the delay time, i.e. the reaction length, $L_\tau = v\tau$, exceeds the characteristic diffusion length, $L_D = \sqrt{2D\tau}$; and (3) long delays randomize dynamics, leading to flat density profiles and zero polarization. These predictions, validated through experiments with thermophoretic microswimmers and Brownian dynamics simulations, establish time delay as a crucial control parameter for particle transport and organization in active matter. Our findings have implications for understanding biological navigation strategies, the evolution of sensory systems, and the design of synthetic microrobots with programmable responses to heterogeneous environments.

\section*{Experimental Setup}

\begin{figure}[h]
\centering
\includegraphics[width=1\columnwidth]{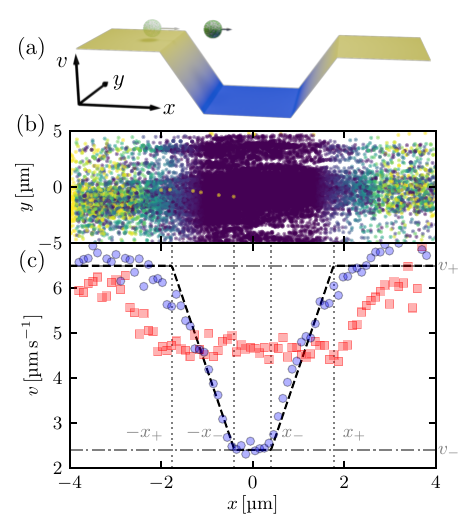}
\caption{{Experimental setup. (a) A spherical thermophoretic microswimmer undergoes active Brownian motion in a spatially-varying laser intensity profile that controls the self-thermophoretic propulsion of the swimmer using a feedback loop (see SI). (b) Sample trajectory of the microswimmer over 15 minutes in an $8\,\si{\micro\metre} \times 10\,\si{\micro\metre}$ chamber, exhibiting rotational diffusion $D_\theta = 2.9\,\Drunit$, translational diffusion $D = 0.04\,\Dunit$, and delay time $\tau = 0.46\,\si{\second}$. Colors indicate instantaneous velocity. (c) Velocity profiles induced by the spatially varying laser intensity: imposed profile with slope $m = \pm3 \, \text{s}^{-1}$ (dashed line), instantaneous velocity profile measured at $\tau = 0.46\,\si{\second}$ (squares), and imposed profile reconstructed from trajectory measurements (circles).}
}
\label{Fig:1}
\end{figure}

Our experimental system employs active particles consisting of melamine resin spheres (diameter $d=2.19\,\si{\micro\metre}$) partially coated with gold nanoparticles ($d_\text{Au}=10\,\si{\nano\metre}$, surface coverage 10\%) suspended in water between parallel glass plates separated by $h=4\,\si{\micro\metre}$ [Fig.~\ref{Fig:1}(a)]. The particles were propelled via self-thermophoresis induced by focusing a $532\,\si{\nano\metre}$ laser ($w_0=0.5\,\si{\micro\metre}$) to the edge of the particle (displaced by distance $d$ from the center of the particle \cite{Franzl2021}), where plasmonic heating generates local temperature gradients and thermo-osmotic flows that propel the particle~\cite{Wuerger2013,Cichos2016}.

The particles were tracked using dark-field microscopy with LED illumination and a high numerical aperture (NA) oil-immersion darkfield condenser (1.2NA) integrated into an Olympus IX-71 microscope. Scattered light was collected through an Olympus 100x/1.35NA objective (aperture set to 0.6NA) and focused onto an Andor iXon EMCCD camera, which imaged a $50\times50\,\si{\micro\metre}^2$ field of view at 20 frames per second. Particle positions were analyzed at video rate during the experiment to control the laser position and power to adjust the particle speed with an acousto-optic deflector system (AA Optics). The feedback control loop has an intrinsic delay of $\tau_0=0.046\,\si{\second}$. To explore the influence of time delay, we introduce an additional programmed delay, obtaining delay times of $\tau=n\tau_0$ ($n=1,4,7,10,13$) such that the current particle speed depends on an earlier position.

We imposed a spatially varying velocity profile \( v(x) \) composed of three regions [Fig.~\ref{Fig:1}(c)]: a low-speed central region (\( |x| \leq x_- = x_m - 2v_-/m \)) with \( x_m = \SI{2}{\micro \metre} \) and constant velocity \( v_- = 2.4\,\Vunit \); transition regions (\( x_- < |x| \leq x_+ = x_m + (v_+ - 3v_-)/m \)), where the velocity increases linearly with slope \( m \) (the rate of change of velocity with position), from \( v_- \) at \( |x| = x_- \) to \( v_+ = 6.5\,\Vunit \) at \( |x| = x_+ \); and high-speed outer regions (\( |x| > x_+ \)) with constant velocity \( v_+ \). In symbols,
\begin{equation}
v(x) = v_- I_{0\, x_-} + \left[3v_- + m(|x| - x_m)\right] I_{x_- x_+} + v_+ I_{x_+ \infty},
\end{equation}
where \( I_{ab}(x) \) is the indicator function, equal to 1 for \( |x| \in [a,b] \) and 0 otherwise, effectively selecting the relevant spatial regions.

We aimed the laser either to the left or right of the particle’s center along the $x$-axis, causing the microspheres to persistently move along the horizontal axis besides their Brownian motion [Fig.~\ref{Fig:1}(b)]. The microspheres underwent rotational Brownian motion, but since the laser focus remained fixed and the heated gold nanoparticles experienced almost instantaneous temperature relaxation, this motion didn’t affect the temperature field or the propulsion direction. To simulate the rotational diffusion inherent in microorganisms, we introduced an artificial angular variable $\theta(t)$ through Brownian dynamics simulation of Eq.~\eqref{eq:theta} to modulate the velocity as $v(x(t-\tau))\cos(\theta(t))$~\cite{SI}. Throughout all experiments, the rotational diffusion coefficient was fixed at $D_\theta=2.9\,\Drunit$.

Figure~\ref{Fig:1}(b) illustrates an example of the trajectory observed for a delay $\tau = 10 \tau_0$ and a speed profile with a slope of $m = 3 \, \text{s}^{-1}$, as depicted by the dashed black line in Fig.~\ref{Fig:1}(c). The delayed response of microswimmers to the velocity profile leads to a mixing of high and low activity regions. Consequently, the mean instantaneous velocity profile $\langle \delta(x(t) - x) \left|\boldsymbol{x}(t+\tau_0) - \boldsymbol{x}(t)\right|/\tau_0 \rangle$ experienced by the microswimmer (red squares) deviates from the imposed one. The imposed velocity profile can be reconstructed from the trajectories as $\langle \delta(x(t-\tau) - x)\left|\boldsymbol{x}(t+\tau_0) - \boldsymbol{x}(t)\right|/\tau_0 \rangle$ (blue circles).

\section*{Results}

To describe the delayed adaptation of microswimmers to environmental changes, we employ an active Brownian particle model incorporating time-delayed speed adjustments. The dynamics are governed by the following coupled stochastic equations (which track how position and orientation evolve over time):
    \begin{eqnarray}
    \dot{\mathbf{x}}(t) &=& v[x(t-\tau)]\mathbf{n}(t) + \sqrt{2D}\boldsymbol{\xi}(t),
    \label{eq:x}\\
    \dot{\theta}(t) &=& \sqrt{2D_\theta}\xi_\theta(t)
    \label{eq:theta}
    \end{eqnarray}
Here, $\mathbf{x}(t) = [x(t),y(t)]$ represents the position and $\mathbf{n}(t) = [\cos \theta(t),\sin \theta(t)]$ the orientation of the microswimmer. The speed $v[x(t-\tau)]$ depends on the particle's position at an earlier time $t-\tau$, while $D$ and $D_\theta$ denote translational and rotational diffusivities. The terms $\boldsymbol{\xi}$ and $\xi_\theta$ represent independent, normalized, unbiased Gaussian white noises.

\begin{figure}
\centering
\includegraphics[width=1\columnwidth]{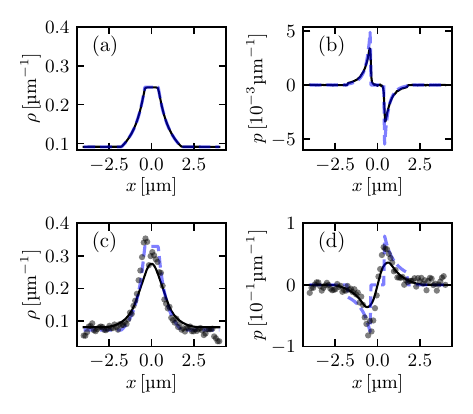}
\caption{Effect of short time delays on density [(a) and (c)] and polarization [(b) and (d)] profiles for the velocity profile in Fig.~\ref{Fig:1}. Results are shown for $D_\theta = 2.9\,\Drunit$, $D = 0.04\,\Dunit$, and $\tau=0\,\si{\second}$ (top) or $\tau=0.184\,\si{\second}$ (bottom). Solid black: simulations; blue dashed: theory [Eqs.~\eqref{eq:rappMR},\eqref{eq:pappMR}]; black circles: experiments.
}
\label{Fig:2}
\end{figure}

\begin{figure}
\centering
\includegraphics[width=1\columnwidth]{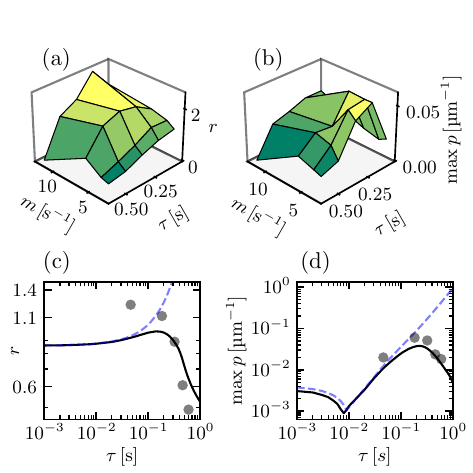}
\caption{Effect of longer delays on density and polarization for $D_\theta = 2.9\,\Drunit$, $D = 0.04\,\Dunit$. Each data point in the first row corresponds to a single measurement with a given delay $\tau$ and slope $m$ of the velocity profile. (a) Experimentally obtained ratio $r$ of the weights of trajectory points inside the region from $-2 \, \rm{\mu m}$ to $2 \, \rm{\mu m}$ versus those outside. (b) Corresponding maximum polarization. (c) and (d) show the density ratio $r$ and maximum polarization as functions of delay for the velocity profile in Fig.~\ref{Fig:1}c. Circles: experiment; blue dashed lines: theoretical predictions from Eqs.~\eqref{eq:rappMR} and \eqref{eq:pappMR}; black solid lines: Brownian dynamics simulations.
}
\label{Fig:3}
\end{figure}

\begin{figure}
\centering
\includegraphics[width=1\columnwidth]{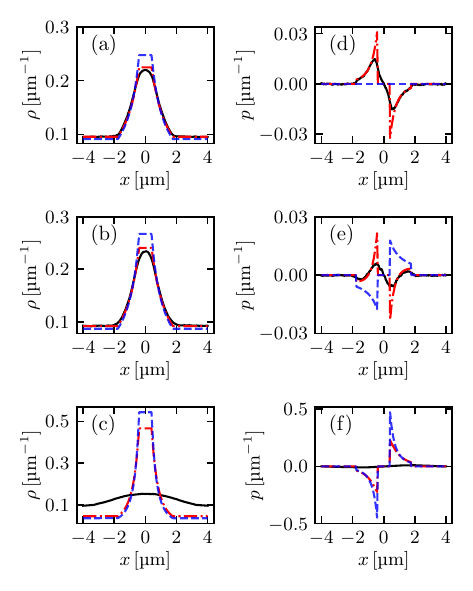}
\caption{Effects of translational diffusion on density and polarization. Data were obtained for $D_\theta = 2.9 \, \rm{s}^{-1}$, the velocity profile in Fig.~\ref{Fig:1}, and a translational diffusion coefficient $D=0.4 \, \Dunit$, which is ten times larger than in experiments.
The individual rows correspond to time delays $\tau = 0, 0.046$, and $0.598\,\rm{s}$, respectively.
Black solid lines: Brownian dynamics simulations; blue dashed and red dot-dashed lines: theoretical predictions from Eqs.~\eqref{eq:rappMR} and \eqref{eq:pappMR} for $D = 0  \, \Dunit$ and $D=0.4 \,\Dunit$, respectively.
 }
\label{Fig:4}
\end{figure}

For \emph{short delay times} (where $\tau \ll D/v^2, 1/D_\theta$), we have derived analytical expressions describing two key properties of the system: the marginal density $\rho$ (the probability of finding a particle at position $x$), and polarization $p$ (the average orientation at position $x$, with positive values indicating rightward bias and negative values indicating leftward bias)~\footnote{$\rho = \langle \delta[x(t)-x] \rangle$ and $p = \langle \cos(\theta(t)) \delta[x(t)-x] \rangle$. The averages are computed over long stationary trajectories of the process $\{x(t), \theta(t)\}$.}, which read

    \begin{eqnarray}
    \rho &=& \rho_0 \left(\frac{1}{v^2 + 2 D D_\theta}\right)^\frac{1+D_\theta\tau}{2}, \label{eq:rappMR} \\
    p &=& \frac{v^2\tau^2 - 2D\tau}{v^2+2 D D_\theta}\frac{\rho}{2\tau}v',\label{eq:pappMR}
    \end{eqnarray}
where $\rho_0$ normalizes the density profile and $v'=\frac{dv}{dx}$. Equations \eqref{eq:rappMR} and \eqref{eq:pappMR} are derived in the Appendix. They can be verified in Brownian dynamics simulations (see Figs.~\ref{Fig:2}--\ref{Fig:4}) and also directly in comparison to experiments (Fig.~\ref{Fig:2}c,d).

Figure \ref{Fig:2}a,b shows the density and polarization profiles for $\tau = 0$, which is not accessible in current experiments but was realized in Ref.~\cite{Sokker2021}. Comparing the graphs in Fig.~\ref{Fig:2}c,d with those in Fig.~\ref{Fig:2}a,b reveals that adding delay enhances the maximum density and, more strikingly, reverses and drastically enhances the polarization. All these effects are captured by Eqs.~\eqref{eq:rappMR} and \eqref{eq:pappMR}. The density enhancement follows from the exponent $1/2 + D_\theta \tau/2 < 1$ in Eq.~\eqref{eq:rappMR}~\cite{Volpe2016,Volpe2018}. According to Eq.~\eqref{eq:pappMR}, the polarization is controlled by the competition between reaction length ($L_\tau = v\tau$) and the characteristic diffusion length ($L_D = \sqrt{2D\tau}$). When $L_D > L_\tau$, the polarization is positive as activity decreases ($v' < 0$), matching previous findings without delay~\cite{Sokker2021,Auschra2021,AuschraCurved2021}. As $\tau$ increases, the polarization decreases, vanishes when the two lengths match, flips sign when the delay exceeds a critical value ($\tau > 2D/v^2$), and subsequently, its absolute value increases almost linearly with $\tau$. These predicted behaviors are verified by simulations in Fig.~\ref{Fig:3}c-d.

In contrast, \emph{long delay times} randomize the dynamics, producing flat density profiles ($\rho \sim 1$) and zero polarization ($p = 0$). These long delay regimes can be characterized by $\tau \gg \tau_c$, where $\tau_c \sim \max(1/D_\theta, L^2/D)$ represents a critical delay time beyond which the particle's motion becomes decorrelated from its position at the time of sensing. Here, $L$ is the characteristic length scale of the velocity gradient.
The transition between short and long delay regimes thus necessarily features peaks in the maximum density and polarization. These predictions are verified in Fig.~\ref{Fig:3}. Specifically, we depict the maximum polarization, $\max_x p$, and the localization in the density measured by the parameter $r = \langle I_{0 2}[x(t)] \rangle / \langle I_{2 \infty}[x(t)] \rangle$, which represents the ratio of the weights of trajectories in the central and outer regions of the arena (with higher values indicating stronger particle accumulation in the central region). Figures~\ref{Fig:3}a,b show $r$ and $\max_x p$ obtained from experiments for various delays, $\tau$, and speed profile slopes, $m$. Figures~\ref{Fig:3}c,d present the same quantities obtained from simulations (solid black lines), experiments (circles), and analytical calculations using Eqs.~\eqref{eq:rappMR} and \eqref{eq:pappMR} (dashed blue lines) for $m = 3 \, \text{s}^{-1}$.
Finally, in Fig.~\ref{Fig:4}, we illustrate the interplay of delay, $\tau$, and translation diffusivity, $D$, on the density and polarization profiles.
Since $D$ is fixed at a small value of $D=0.04\,\Dunit$ in the experiments, we present results for a tenfold larger translational diffusivity, $D=0.4 \,\Dunit$, obtained from simulations (solid black lines) and theoretical predictions~\eqref{eq:rappMR} and \eqref{eq:pappMR} (red dot-dashed lines). Neglecting $D$ in the theoretical predictions (dashed blue lines) systematically underestimates the maximum density for short delays (Fig.~\ref{Fig:4}a,b). Moreover, for $\tau = 0$, Eq.~\eqref{eq:pappMR} incorrectly predicts vanishing polarization (Fig.~\ref{Fig:4}d), and for $\tau = \tau_0$, it produces a polarization profile with the wrong sign (Fig.~\ref{Fig:4}e). For a long delay $\tau = 13\tau_0$, the short-delay theory fails to predict both density and polarization, regardless of $D$ (Fig.~\ref{Fig:4}c,f).

In summary, we theoretically predicted that microswimmers sensing an activity profile with a delay exhibit maxima in both localization and polarization as functions of the delay time, and demonstrated the phenomenon of delay-induced polarization inversion. These predictions were validated through Brownian dynamics simulations and confirmed experimentally using thermophoretic microswimmers. Our experimental realization, the first of its kind at the microscale in the low Reynolds number regime, enables precise control of activity-rotational diffusivity landscapes and interaction delays. Our findings establish time delay as a fundamental control parameter in active matter systems, with the sign inversion of polarization offering a novel mechanism for directing microscale transport without external fields~\cite{rhein2025}. This principle extends beyond our experimental system to diverse biological contexts, potentially explaining how microorganisms optimize their response times for competitive advantages in heterogeneous environments. Future research could leverage these insights to design adaptive microrobotic systems with programmable delays, creating self-organizing swarms capable of navigating complex landscapes for targeted medical interventions. Moreover, the demonstrated interplay between delay time and diffusion opens new avenues for understanding cellular decision-making processes, where sensory delays may serve not as limitations but as evolutionary advantages that filter environmental noise and enhance collective behavior. Ultimately, this work contributes to a broader paradigm shift in non-equilibrium physics, where temporal programming becomes as important as spatial patterning in controlling the emergent properties of active matter.

\section{Appendix: Theory}
Let us now sketch the derivation of our main analytical results. For more detailed derivations, see the SI~\cite{SI}. The joint probability density $\hat{\rho}(x,\theta,t)$ for position $x$ and angle $\theta$ in Eqs.~\eqref{eq:x} and \eqref{eq:theta} obeys the Fokker-Planck equation~\cite{Guillouzic1999,loos2021stochastic}
\begin{multline}
    \partial_t \hat{\rho}(x,\theta,t) = (D\partial^2_x + D_\theta \partial^2_\theta) \hat{\rho}(x,\theta,t)
    \\- \partial_x \int dx' \int d \theta'
    v(x')\cos\theta \hat{P}(x,\theta,t;x',\theta',t-\tau).
    \label{eq:FPEfull}
\end{multline}
This equation is not closed due to the presence of two-time PDF $\hat{P}(x,\theta,t;x',\theta',t')$ on the right-hand side, and an exact closure is known for linear functions $v(x)$ only~\cite{loos2021stochastic}.

We are interested in the stationary density $\rho(x) = \lim_{t\to \infty} \int d\theta \hat{\rho}(x,\theta,t)$ and polarization $p(x) = \lim_{t\to \infty} \int d\theta \cos \theta\hat{\rho}(x,\theta,t)$. Assuming stationarity $\partial_t \hat{\rho}(x,\theta,t) = 0$ and that the current $j_x = - D\partial_x \rho(x) + \int dx' \int d \theta' v(x')\cos\theta \hat{P}(x,\theta,t;x',\theta',t-\tau)$ vanishes due to the symmetry, we find the partial differential equations~\cite{SI}
\begin{multline}
    D \partial_x \rho(x) = \\
    \int d\theta \int dx' \int d\theta' v(x')\cos\theta \hat{P}(x,\theta,t;x',\theta',t-\tau),
    \label{eq:rhoF}
\end{multline}
\begin{multline}
    D \partial^2_x p(x) - D_\theta p(x) =
    \\ \partial_x \int d\theta \int dx' \int d \theta' v(x')\cos^2\theta \hat{P}(x,\theta,t;x',\theta',t-\tau).
    \label{eq:pF}
\end{multline}
 To solve them, we need to approximate the unknown two-time propagator $\hat{P}(x,\theta,t;x',\theta',t-\tau)$.

 For long delays, $\tau \gg D/v^2, 1/D_\theta$, it is reasonable to assume that
 that positions $x$ and $x'$ and orientations $\theta$ and $\theta'$ in the two-time PDF $\hat{P}(x,\theta,t;x',\theta',t-\tau)$ are independent, i.e., $\hat{P}(x,\theta,t;x',\theta',t-\tau) \approx \hat\rho(x,\theta) \hat\rho(x',\theta')$,  where $\hat\rho(x,\theta) = \lim_{t\to \infty}\hat\rho(x,\theta,t)$. Substituting this approximation in Eqs.~\eqref{eq:rhoF}--\eqref{eq:pF} yields
 \begin{eqnarray}
     D \rho' &= \left<v\right> p,\\
     D p'' - D_\theta p &=
    \left<v\right> \left<\cos^2\theta \right>' \approx 0,
 \end{eqnarray}
 with $\left<v\right> = \int dx v(x)\rho(x)$ and $f' = \partial_x f(x)$. The term $\left<\cos^2\theta \right> = \int d\theta \cos^2 \theta \hat\rho(x,\theta)$ and we approximate it by $1/2$, which corresponds to $\rho(x,\theta) \approx \rho(x)/2\pi$ and is a good approximation for vanishing delay~\cite{Sokker2021,Auschra2021}. The only solution to these equations that obeys the boundary conditions of the problem is a flat distribution $\rho(x) = const.$ and vanishing polarization $p(x) = 0$~\cite{SI}.

 For short delay times, $\tau \ll D/v^2, 1/D_\theta$, one might use the Gaussian short-time approximation of the Fokker-Planck propagator~\cite{risken1996fokker}. However, closed results can only be obtained after further approximating the Gaussian by a $\delta$-function: $\hat{P}(x,\theta,t;x',\theta',t-\tau) \approx
 \delta(\theta-\theta') \delta(x'- x + \tau v(x)\cos\theta)  \rho(x,\theta)$, which can be done when $\tau D_\theta \to 0$ and $\tau D \to 0$. Using this approximation, again assuming that $\langle\cos^2\theta \rangle \approx 1/2$, and expanding everything up to the leading order in $\tau$ yields the equations
\begin{eqnarray}
p'' &=&  \left(\frac{v^2}{2 D^2} + \frac{D_\theta}{D}\right) p + \frac{v'}{2 D}\left(1 - \tau\frac{v^2}{2 D} \right)\rho,\label{eq:p}\\
D \rho' &=& v p - \frac{\tau}{2} v v' \rho.
\label{eq:rho}
\end{eqnarray}
They generalize the equations in Refs.~\cite{Sokker2021,Auschra2021} for zero delay. The system can be solved exactly for a piece-wise constant velocity profile~\cite{SI} and is difficult to tackle otherwise. For smooth enough profiles, one can neglect the term $p''$, which yields the solution in Eqs.~\eqref{eq:rappMR} and \eqref{eq:pappMR}. While much simpler, our derivation generalizes that in Refs.~\cite{Volpe2016,Volpe2018} by taking into account the translational diffusivity and also by predicting the polarization.

\section*{Acknowledgements}
V.H. acknowledges the support of Charles University through project PRIMUS/22/SCI/009.


\bibliographystyle{apsrev4-1}
\bibliography{references}

\end{document}